\documentstyle{article}
\begin{document}

\title{\bf A note on topological brane theories}
\author{\bf Ingemar Bengtsson\thanks{Email address: ingemar@physto.se}, 
Nuno Barros e S\'a\thanks{Email address: nunosa@physto.se}  and 
Maxim Zabzine\thanks{Email address: zabzin@physto.se}\\
\,\\
$^{*\dagger\ddagger}$Fysikum, Stockholms Universitet, Box 6730,
113 85 Stockholm, Sweden\\
$^\dagger$DCTD, Universidade dos A\c cores, 
9500 Ponta Delgada, A\c cores, Portugal.}
\date{}
\maketitle

\vskip -7.0cm
 \hfill USITP-00-06\\
\vskip 7.0cm

\begin{abstract}
We consider the theory of closed
$p$-branes propagating on $(p+1)$-dimensional space-time
manifolds. This theory has no local degrees of freedom. 
Here we study its canonical and BRST structures
of the theory. In the case of  
locally flat backgrounds one can show that the $p$-brane theory
is related to another known topological field theory. 
In the general situation some equivalent actions 
can also be written for the topological $p$-brane theory. 
\end{abstract}

\section{Introduction}

In the present note we consider a $p$-brane propagating on a $(p+1)$%
-dimensional background manifold. In this model all degrees of freedom can
be gauged away locally. However there may still be non trivial non-local
(topological) degrees of freedom. The motivation for the study of this model
is twofold. First this theory is interesting by itself as a topological
field theory \cite{Birmingham:1991ty}, and we shall see that there is
something one can learn about its canonical and BRST structures. Second the
present theory may serve as a toy model for the study of general extended
objects which play an important role in modern string theory. Thus the
present model can give some insight into the more general case of $p$-branes
propagating on background manifolds of dimension higher than $(p+1)$.

The $p$-brane theory describes the embedding of a $(p+1)$-dimensional
world-volume into a $d$-dimensional space-time manifold ${\cal M}$. The
action is given by the volume of the embedded $(p+1)$-dimensional manifold 
\begin{equation}
S[X]=-T\int d^{p+1}\xi \sqrt{-\det \left( G_{\mu \nu }(X)\partial _{\alpha
}X^{\mu }\partial _{\beta }X^{\nu }\right) }\ ,  \label{s1}
\end{equation}
where $G_{\mu \nu }$ is the metric in the $d$-dimensional space-time
manifold. Throughout the paper we shall look at the case
of Minkowskian signature, and we shall see
that the results can be generalized to the Euclidean case in a
straightforward way. It is assumed that metric $G_{\mu \nu }$ is not
degenerate at any point of manifold ${\cal M}$, $\det (G_{\mu \nu })\neq 0$.
The action (\ref{s1}) is called the Nambu-Goto action. The parameter $T$ is
called tension and it is a direct generalization of the concept of mass to $%
p$-branes. The $p$-brane action can be equivalently written in the Polyakov
form 
\begin{equation}
S[X,h]=-\frac{T}{2}\int d^{p+1}\xi \ \sqrt{-h} \left[ h^{\alpha \beta
}\partial _{\alpha }X^{\mu }\partial _{\beta }X^{\nu }G_{\mu \nu
}(X)- (p-1)\right] \ ,  \label{pol}
\end{equation}
where the $h_{\alpha \beta }$ transform as a world-volume metric ($h\equiv
\det (h_{\alpha \beta })$) and play the role of Lagrange multipliers. The
equivalence can be checked by varying this action with respect to $h_{\alpha
\beta }$, solving the resulting equations of motions and putting the
resulting solution for $h_{\alpha \beta }$ into equation (\ref{pol}). The
result is the action (\ref{s1}). One should notice that within the
equivalence between actions (\ref{s1}) and (\ref{pol}) the induced
world-volume metric must be nondegenerate.

Now let us discuss the symmetries of the theory. There is a local
diffeomorphism invariance for the action (\ref{s1}) 
\begin{equation}
\delta X^{\mu }=\pounds _{\zeta }X^{\mu }\ ,  \label{rep}
\end{equation}
where $\pounds _{\zeta }$ stands for the Lie derivative along $\zeta
^{\alpha }$. For the Polyakov action (\ref{pol}) the transformation (\ref
{rep}) should be supplemented by the appropriate transformation for the
auxiliary world-volume metric $\delta h^{\alpha \beta }=\pounds _{\zeta
}h^{\alpha \beta }$. For the case $p=1$ there is an extra local symmetry $%
\delta h^{\alpha \beta }=\Lambda h^{\alpha \beta }$ (Weyl rescaling). In
addition both actions (\ref{s1}) and (\ref{pol}) are invariant under
arbitrary diffeomorphisms on ${\cal M}$, if $G_{\mu \nu }$ is transformed
properly.

The local symmetry (\ref{rep}) allows one to choose locally the following
gauge 
\begin{equation}
X^{\mu }=\xi ^{\mu }\ ,\ \mu =0,1,...,p\ ,  \label{static}
\end{equation}
which is usually called the static gauge. The existence of static gauge 
 can be argued from the picture of embedding of a manifold into another one.
 Thus in the case of interest $d=p+1$
one can locally gauge away all degrees of freedom. However on a nontrivial
background manifold ${\cal M}$ one cannot do this globally and therefore
there are nontrivial global (topological) degrees of freedom. One can also
see that degenerate situations (when $\det \left( G_{\mu \nu }(X)\partial
_{\alpha }X^{\mu }\partial _{\beta }X^{\nu }\right) =0$) do not appear 
 because of (\ref{static}). 
 In the static gauge the determinant of the induced metric is
equal the to the determinant of the background metric which is assumed to be
nondegenerate. Therefore as soon as we want to keep the local diffeomorphism
symmetry (\ref{rep}) (i.e. the picture of embedding of one manifold into
 another) it is assumed everywhere that 
\begin{equation}
\det \left( G_{\mu \nu }(X)\partial _{\alpha }X^{\mu }\partial _{\beta
}X^{\nu }\right) \neq 0\ .
\end{equation}
Throughout the paper we use the following notation: $\mu $, $\nu $ denote
space-time indices, $\alpha $, $\beta $ world-volume indices and $a$, $b$, $c$
spatial world-volume indices.

Let us assume that the space-time manifold ${\cal M}$ is compact and
oriented. When we come to the Hamiltonian treatment we also assume that $%
{\cal M}=R\times \Sigma $ where $\Sigma $ is a compact and oriented spatial
manifold. $p$-branes can be closed or open. For closed $p$-branes
periodicity conditions must be imposed along all spatial directions. The
analysis of open $p$-branes is more involved since the theory should be
supplemented by appropriate boundary conditions. 
 In this note we look only at
closed $p$-branes. In the case of closed branes the Nambu-Goto action is a
constant for all field configurations and it is equal to the volume of the
background manifold ${\cal M}$. We hope to come back to 
 the case of open $p$%
-branes elsewhere.

In this paper we study mainly the classical aspects of the theory. The paper
is organized as follows: In section 2 we go through the Hamiltonian
treatment of the closed brane theory. Three equivalent sets of constraints
are presented. In section 3 we take a look at the construction of BRST
generators for these three sets of constraints. Three different BRST
generators are related to each other through canonical transformations in
the extended phase space. In section 4 we look at equivalent forms of the
action and specifically discuss the case of locally flat backgrounds. The
degrees of freedom are briefly considered and the subtleties related to
 the degenerate solutions are pointed out. In the last section we discuss
the results and outline possible generalizations of the model.

\section{Hamiltonian treatment}

In this section we take a look at the Hamiltonian treatment of the
system. The $p$-brane theory is a generally covariant system and therefore
the naive Hamiltonian vanishes identically. In this theory 
the full Hamiltonian is given by a linear combination of the corresponding
constraints which are first class. 
 Our goal is to write down three different sets of constraints
 for the model .

In order to carry out the Hamiltonian formulation of the theory we choose
one of the integration variables $\xi ^{\alpha }$ as the evolution parameter
(in the case of a relativistic metric with signature $(-1,1,...,1)$, the
one-parameter group of diffeomorphisms defined by the translations in that
variable should be generated by a timelike vector field), which we take to
be $\xi ^{0}$; the remaining integration variables, which parametrize the
brane itself, are represented by $\xi ^{a}$, with the small Latin letters
taking values from $1$ to $p$. The system is totally constrained since
the theory is invariant under redefinitions of the evolution parameter.

Denoting by $P_{\mu }$ the momenta conjugate to the $X^{\mu }$ and starting
from either the Polyakov action (\ref{pol}) or the Nambu-Goto action (\ref
{s1}) the constraints can be worked out as \cite{c1} 
\begin{equation}
{\cal H}_{I}=\left( 
\begin{array}{c}
{\cal H} \\ 
{\cal H}_{a}
\end{array}
\right) =\left( 
\begin{array}{c}
G^{\mu \nu }(X)P_{\mu }P_{\nu }+T^{2}\det [q_{ab}]\  \\ 
P_{\mu }\partial _{a}X^{\mu }
\end{array}
\right) \ ,  \label{co2}
\end{equation}
where 
\begin{equation}
q_{ab}=G_{\mu \nu }(X)\partial _{a}X^{\mu }\partial _{b}X^{\nu }
\end{equation}
is the induced spatial metric on the brane. The constraints (\ref{co2}) are
first class and obey the algebra 
\begin{eqnarray}
\left\{ {\cal H}_{a}[M^{a}],{\cal H}_{b}[N^{b}]\right\} &=&{\cal H}%
_{a}[\pounds _{M}N^{a}]\quad ,  \label{f1} \\
\left\{ {\cal H}_{a}[M^{a}],{\cal H}[N]\right\} &=&{\cal H}[\pounds _{M}N]
 \quad ,
\label{f2} \\
\left\{ {\cal H}[M],{\cal H}[N]\right\} &=&{\cal H}_{a}[q q^{ab}(M\partial
_{b}N-N\partial _{b}M)]\quad ,  \label{f3}
\end{eqnarray}
where $\pounds _{N}$ stands for the Lie derivative along the vector field $%
N^{a}$, and $q=\det \left( q_{ab}\right) $. Since there are $d$ pairs of
canonical conjugate variables and $p+1$ constraints, the theory possesses $%
(d-p-1)$ degrees of freedom per brane point. Therefore in the case of a $%
(d-1)$-brane one has got no dynamical degrees of freedom and the theory is
purely topological. The algebra (\ref{f1})-(\ref{f3}) is called the algebra
of many-fingered time (the name is due to Wheeler). The constraints (\ref
{co2}) and their algebra (\ref{f1})-(\ref{f3}) are true for a $p$-brane in
any space-time dimension $d$. The algebra (\ref{f1})-(\ref{f3}) is closed
only for the case $p<2$. Now let us analyze the specific properties for a $p$%
-brane propagating on a $(p+1)$-dimensional space-time.

Starting from the Nambu-Goto action (\ref{s1}) one can see that the
constraints can be written in a form in which all of them are linear in the
momenta. In order to do so, we observe that for $p$-branes the dimension of
the world-volume in equation (\ref{s1}) is the same as the dimension of the
embedding space-time, namely $p+1$. Consequently $\partial _{\alpha }X^{\mu
} $ is a square matrix, and one can write 
\begin{equation}
\det \left( G_{\mu \nu }(X)\partial _{\alpha }X^{\mu }\partial _{\beta
}X^{\nu }\right) =G\left( X\right) \det {^{2}}\left( \partial _{\alpha
}X^{\mu }\right) \ ,
\end{equation}
with 
\begin{equation}
G\left( X\right) =\det \left( G_{\mu \nu }(X)\right) \ .
\end{equation}
The action (\ref{s1}) becomes then 
\begin{equation}
S=\pm T \int d^{p+1}\xi \sqrt{-{\cal G}}\frac{1}{(p+1)!} 
\epsilon ^{\alpha _{0}...\alpha _{p}}\epsilon _{\mu _{0}...\mu
_{p}}\partial _{\alpha _{0}}X^{\mu _{0}}...\partial _{\alpha _{p}}X^{\mu
_{p}}\ ,  \label{linact}
\end{equation}
where $\pm $ corresponds to the two possible solutions of the square root.
Let us keep both signs in all calculations and eventually one can see that
the sign ambiguity corresponds to the two possible orientations on the
manifold. The equations of motion for (\ref{linact}) are somewhat trivial.
They tell us that the exterior derivative of the volume form is zero. The $%
(p+1)$ decomposition of (\ref{s1}) is straightforward, 
\begin{equation}
S=\pm T\int d\xi ^{0}d^{p}\xi \ n_{\mu }\dot{X}^{\mu }\ ,
\end{equation}
where a dotted quantity represents its derivative with respect to the
evolution parameter $\xi ^{0}$, and the vector $n_{\mu }$ is a function of
the configuration variables $X^{\mu }$ given by 
\begin{equation}
n_{\mu }=\frac{1}{p!}\sqrt{-G}\epsilon _{\mu \nu _{1}...\nu _{p}}\epsilon
^{a_{1}...a_{p}}\partial _{a_{1}}X^{\nu _{1}}...\partial _{a_{p}}X^{\nu _{p}}%
\ .
\end{equation}
The vector $n_{\mu }$ satisfies the following properties 
\begin{eqnarray}
G^{\mu \nu }n_{\mu }n_{\nu } &=&-\det [q_{ab}]\ , \\
n_{\mu }\partial _{a}X^{\mu } &=&0\ .
\end{eqnarray}
In this form it is clear that the momenta conjugate to the $X^{\mu }$ are 
\begin{equation}
P_{\mu }=\pm Tn_{\mu }\ .  \label{c2}
\end{equation}
The Hamiltonian vanishes and one must include the primary constraints given
by equation (\ref{c2}) with the aid of some Lagrange multiplier functions $%
\lambda ^{\mu }$, 
\begin{eqnarray}
L^{\pm } &=&\int d^{p}\xi \ \left[ P_{\mu }\dot{X}^{\mu }-\lambda ^{\mu }%
{\cal C}_{\mu }^{\pm }\right] \ , \\
{\cal C}_{\mu }^{\pm } &=&P_{\mu }\mp Tn_{\mu }(X)\ .  \label{s2}
\end{eqnarray}
The obtained constraints are linear in the momenta and its Poisson bracket
algebra vanishes strongly. The two sets of constraints ${\cal C}_{\mu }^{+}$
and ${\cal C}_{\mu }^{-}$ correspond to two different branches of the
constraint surface. These two sets intersect only on the degenerate
solutions 
\begin{equation}
{\cal C}_{\mu }^{+}={\cal C}_{\mu }^{-}=0\quad \Longrightarrow \quad n_{\mu
}=0\quad \Longrightarrow \quad \det \left( q_{ab}\right) =0\ .
\end{equation}
Thus exluding degenerate solutions one has two 
 independent branches of the
theory with different Lagragians $L^{\mp }$. The constraints ${\cal C}_{\mu
}^{\pm }$ generate the following transformation 
\begin{equation}
\delta X^{\mu }=\{X^{\mu },{\cal C}_{\mu }^{\pm }[N^{I}]\}=N^{\mu }\ .
\label{tra}
\end{equation}
In fact one can see that this is the real symmetry of the action (\ref
{linact}). The natural question might arise about the relation between (\ref
{tra}) and the local diffeomorphism invariance (\ref{rep}). There is a
one to one map between the two transformations 
\begin{equation}
\delta X^{\mu }=\pounds _{\zeta }X^{\mu }=(\partial _{\alpha }X^{\mu })\zeta
^{\alpha }=N^{\mu }\ ,
\end{equation}
 when the quadratic matrix $(\partial _{\alpha }X^{\mu })$ is assumed to
  be nondegenerate. 
 Thus for every vector $\zeta $ there is unique vector $N$ and
vice versa. However it should be stressed that in general the gauge 
 symmetries (\ref{rep}) and (\ref{tra}) have different properties.
  By using transformation (\ref{tra}) one can bring a nondegenerate
 solution to a degenerate one. One cannot do this by using the 
 transformation (\ref{rep}).  

To get another set of constraints one can contract ${\cal C}_{\mu }^{\pm }$
with the set of independent vectors $G^{\mu \nu }n_{\nu }$ and $\partial
_{a}X^{\mu }$, resulting respectively in 
\begin{equation}
{\cal H}_{I}^{\pm }=\left( 
\begin{array}{c}
{\cal H}^{\pm } \\ 
{\cal H}_{a}
\end{array}
\right) =\left( 
\begin{array}{c}
G^{\mu \nu }n_{\mu }P_{\nu }\pm T\det [q_{ab}] \\ 
P_{\mu }\partial _{a}X^{\mu }
\end{array}
\right)  \ . \label{ca2}
\end{equation}
Comparing with equations (\ref{co2}) we see that only the scalar constraint
is modified, now being linear in the momenta, not quadratic. The constraints
(\ref{ca2}) obey the same Poisson bracket algebra as (\ref{f1})-(\ref{f3}).
Again the two sets of constraints ${\cal H}_{I}^{+}$ and ${\cal H}_{I}^{-}$
describe the two separate branches if the degenerate solutions are excluded,
and they obey two many-fingered algebras on the corresponding independent
branches of the theory. The constraints (\ref{ca2}) basically tell us that
the system can be thought as a parametrized field theory (parametrized
cosmological constant term) \cite{Kuchar:1980ht}. The equation of motion for 
$X^{\mu }$ is given by 
\begin{equation}
\dot{X}^{\mu } =\{X^{\mu },{\cal H}_{a}[N^{a}]+{\cal H}^{\pm }[M]\}= 
N^{a}\partial _{a}X^{\mu }+MG^{\mu \nu }n_{\nu } \ ,
\end{equation}
which is nothing else but the geometrodynamical canonical decomposition \cite
{Kuchar:1980ht} with respect to basic vectors ($G^{\mu \nu }n_{\nu }$, $%
\partial _{a}X^{\mu }$).

Now one can check explicitly the equivalence of these three sets of
constraints, ${\cal C}_{\mu }^{\pm }$, ${\cal H}_{I}^{\pm }$ and ${\cal H}%
_{I}$. It is clear that ${\cal C}_{\mu }^{\pm }$ implies both ${\cal H}%
_{I}^{\pm }$ and ${\cal H}_{I}$. To check the converse we note that the
second equations ${\cal H}_{a}=0$ in the sets of constraints, (\ref{co2})
and (\ref{ca2}), have the general solution 
\begin{equation}
P_{\mu }=\alpha n_{\mu }\ ,
\end{equation}
where $\alpha $ is any function which does not carry indices. Plugging this
result into the last equation of ${\cal H}$ one gets 
\begin{eqnarray}
&&(T^{2}-\alpha ^{2})\det \left( q_{ab}\right) =0\,\,\quad \Longrightarrow 
\nonumber \\
&\Longrightarrow &\quad T=\alpha\ \ \rm{or}\ \ T=-\alpha\ \ \rm{or}\ \ 
\det [q_{ab}]=0\ .
\end{eqnarray}
The first solution is just ${\cal C}_{\mu }^{+}$, the second is ${\cal C}%
_{\mu }^{-}$ and the last one corresponds to a degenerate solution. Applying
the same procedure to ${\cal H}^{\pm }$ one finds that 
\begin{equation}
{\cal C}_{\mu }^{+}=0\,\,\,\,  \Leftrightarrow \ \,\,\,\,
 {\cal H}_{I}^{+}=0\ ,\ {\cal C}
_{\mu }^{-}=0\,\,\,\,  \Leftrightarrow\,\,\,\,  {\cal H}_{I}^{-}=0\ ,
\end{equation}
if $\det \left( q_{ab}\right) \neq 0$ is assumed.

Thus we have shown that all three sets of constraints are equivalent 
\begin{equation}
{\cal C}_{\mu }^{+}=0\,\,\,\, \rm{ and }\,\,\,\, 
{\cal C}_{\mu }^{-}=0\quad \Leftrightarrow
\quad {\cal H}_{I}=0\quad \Leftrightarrow \quad 
{\cal H}_{I}^{+}=0\,\,\,\, \rm{ and }\,\,\,\, 
{\cal H}_{I}^{-}=0 \ ,
\end{equation}
if the case of degenerate metric is excluded and therefore these three sets
of constraints describe the same constrained surface. Since the manifold $%
\Sigma $ is assumed oriented the two branches of the theory are dynamically
independent.

\section{BRST generators}

In this section we construct the BRST generators which correspond to the
different sets of constraints discussed in the previous section.

We have a first class constrained system and its Hamiltonian is a linear
combination of the constraints $\Psi _{I}$. Introducing the ghost variables $%
\eta ^{I}$ and the ghost momenta ${\cal P}_{I}$ one can define the classical
Grassmann-odd BRST generator (charge) $Q$ in the extended phase space 
\cite{Henneaux:1985kr} 
\begin{eqnarray}
Q &=&\int d^{p}\xi \eta ^{I}(\xi )\Psi _{I}(\xi )+\sum\limits_{n=0}^{r}\int
d^{p}\xi _{1}...\int d^{p}\xi _{n}  \nonumber \\
&&Q^{I_{1}...I_{n}}
(\xi _{1},...,\xi _{n}){\cal P}_{I_{1}}(\xi _{1})...
{\cal P}_{I_{n}}(\xi _{n})\ ,  \label{brst}
\end{eqnarray}
such that $Q$ is nilpotent and real. The ghost number of $\Omega $ should be
equal to $1$. 
 The BRST construction is important because it reveals that the
different representations of the constraint surface can be thought of as being
obtained from each other by a canonical transformation in the extended phase
space. In expression (\ref{brst}) the number $r$ is called the rank of the
BRST generator. The concept of rank is not intrinsic and
can be made equal to zero by appropriate redefinitions of constraints 
 \cite{Henneaux:1985kr}.

For the general case of $p$-branes with constraints ${\cal H}_{I}$ given by (%
\ref{co2}) the classical BRST generator $Q$ has been constructed by Henneaux 
\cite{Henneaux:1983um}. The rank of $Q$ is equal $p$. Now we can construct
the BRST generator for the other two sets of constraints ${\cal C}_{\mu
}^{\pm }$ and ${\cal H}_{I}^{\pm }$. Since the Poisson bracket algebra for $%
{\cal C}_{\mu }^{\pm }$ vanishes strongly (thus the algebra is commutative)
the BRST operator $Q^{\pm }$ has rank 0 
\begin{equation}
Q^{\pm }=\int d^{p}\xi \eta ^{\mu }(\xi ){\cal C}_{\mu }^{\pm }(\xi )\ , 
 \label{Qsim}
\end{equation}
where $Q^{+}$ and $Q^{-}$ are defined for the two different sectors. In this
case the BRST transformation ($\delta _{\pm }A=\{A,Q_{\pm }\}$) in the
extended phase space has a simple form 
\begin{eqnarray}
\delta _{\pm }X^{\mu } &=&\eta ^{\mu }\ , \delta _{\pm }\eta ^{\mu }=0
\ ,\delta _{\pm }{\cal P}_{\mu }=-\Phi _{\mu }\ ,  \nonumber \\
\delta _{\pm }P_{\mu } &=&\mp T\frac{1}{(p-1)!}\sqrt{-G}\epsilon _{\mu \nu
_{1}\nu _{2}...\nu _{p}}\epsilon ^{a_{1}...a_{p}} 
\partial _{a_{1}}\eta ^{\nu _{1}}\partial _{a_{2}}X^{\nu _{2}}...\partial
_{a_{p}}X^{\nu _{p}}.
\end{eqnarray}

The BRST generator ${\cal Q}^{\pm }$
  for the constraints ${\cal H}_{I}^{\pm }$ can also be
worked out, being given by 
\begin{eqnarray}
{\cal Q}^{\pm }&=&\int d^{p}\xi \left[ \eta H^{\pm }+\eta ^{a}H_{a}+\left(
\eta ^{a}\partial _{a}\eta +\eta \partial _{a}\eta ^{a}\right) {\cal P+}
\right.  \nonumber \\
&&\left. +\left( q_{\circ }q^{ab}\eta \partial _{a}\eta +\eta ^{a}\partial
_{a}\eta ^{b}\right) {\cal P}_{b}\right] \ ,
\end{eqnarray}
where $\left( \eta ,{\cal P}\right) $ and $\left( \eta ^{a},{\cal P}%
_{b}\right) $ are the ghost pairs associated with ${\cal H}^{\pm }$ and $%
{\cal H}_{a}$ respectivelly. Its rank is 1. Thus we have constructed three
different BRST generators for the same theory. They should relate to each
other by canonical transformations in the extended phase space. We were
unable to construct these canonical transformations in any simple closed
form. However one can certainly construct them perturbatively in same
fashion as in \cite{Henneaux:1985kr} and \cite{Hwang:1991mx}.

As we saw  the three sets of constraints ${\cal H}_{I}$, ${\cal H}%
_{I}^{\pm }$ and ${\cal C}_{\mu }^{\pm }$ describe the  same constraint
surface if we exclude degenerate solutions. There should be the following
relation among the sets of constraints which describe the same constrain
surface 
\begin{equation}
{\cal H}_{I}=(S^{\pm })_{I}^{\mu }{\cal C}_{\mu }^{\pm }\ \ ,\ \ {\cal H}
_{I}^{\pm }=(S)_{I}^{\mu }{\cal C}_{\mu }^{\pm }\ ,  \label{B1}
\end{equation}
where $(S^{\pm })_{I}^{\mu }$ and $(S)_{I}^{\mu }$ must be non degenerate.
It is not difficult to construct these matrices explicitly. Thus for $S^{\pm
} $ we have the following expression 
\begin{equation}
S^{\pm }=\left( 
\begin{array}{lll}
\partial _{1}X^{0} & ... & \partial _{1}X^{p} \\ 
\partial _{2}X^{0} & ... & \partial _{2}X^{p} \\ 
... & ... & ... \\ 
G^{0\nu }(P_{\nu }\pm Tn_{\nu }) & ... & G^{p\nu }(P_{\nu }\pm Tn_{\nu })
\end{array}
\right) \ ,
\end{equation}
and for $S$ the following 
\begin{equation}
S=\left( 
\begin{array}{llll}
\partial _{1}X^{0} & \partial _{1}X^{1} & ... & \partial _{1}X^{p} \\ 
\partial _{2}X^{0} & \partial _{2}X^{1} & ... & \partial _{2}X^{p} \\ 
... & ... & ... & ... \\ 
G^{0\nu }n_{\nu } & G^{1\nu }n_{\nu } & ... & G^{p\nu }n_{\nu }
\end{array}
\right) \ .
\end{equation}
Let us calculate the determinants of these matrixes 
\begin{eqnarray}
\det (S^{\pm }) &=&(-1)^{p}n_{\mu }G^{\mu \nu }(P_{\nu }\pm Tn_{\nu
})=(-1)^{p}\tilde{{\cal H}}^{\pm } \ ,\\
\det (S) &=&(-1)^{p+1}\det [q_{ab}]\ .
\end{eqnarray}
We see that $S$ is non degenerate if degenerate solutions ($\det
[q_{ab}]=0$) are excluded. The matrix $S^{+}$ is not degenerate either as
long as we stay at the branch defined by ${\cal C}_{\mu }^{+}=0$ (or
equivalently by ${\cal H}_{I}^{+}$) and $S^{-}$ is not degenerate at the
branch defined by ${\cal C}_{\mu }^{-}$ (or equivalently by ${\cal H}%
_{I}^{-} $). Therefore using these matrices one can construct perturbatively
the relevant canonical transformations in the extended phase space.

\section{$p$-brane theory in locally flat background}

To understand the theory better we would like to study alternative
representations of this model. In many cases alternative representations of
a theory may help to analyze their degrees of freedom. In this section we
study some classically equivalent actions and analyze the degrees of freedom
corresponding to the topological $p$-brane theory. It is hard to say
anything explicit about the degrees of freedom when the theory is formulated
in the form (\ref{s1}) or (\ref{pol}). Intuitively we understand that the
number of degrees of freedom is related to the number patches needed to
cover the manifold ${\cal M}$. However it is hard to count them explicitly.
Therefore we can try to reformulate the theory in a more transparent way.
One can reach this goal by using new variables. Since the task is difficult
for generic curved background manifolds, we look first at the case of
locally flat manifolds ${\cal M}$ 
\begin{equation}
G_{\mu \nu }=\eta _{\mu \nu }\ .  \label{G}
\end{equation}
At the end of this section we will take a brief look on equivalent actions
for the generic case. However it is still problematic to analyze the degrees
of freedom in all generality.

Now we are assuming that (\ref{G}) holds. Let us enlarge the gauge symmetry
of the system defining the tetrad fields 
\begin{equation}
e_{a}{^{\mu }}=\partial _{a}X^{\mu }
\end{equation}
as the new configuration variables. They are subject to the constraints 
\begin{equation}
\partial _{\lbrack a}e_{b]}{^{\mu }}=0\ .  \label{noco}
\end{equation}
One can easily see that there is a one to one correspondence between new and
old variables in the locally flat space-time. The static gauge (\ref{static}%
) in new variables corresponds to $e_{a}{^{\mu }}=\delta _{a}{^{\mu }}$.

The action in these new variables and their canonical conjugate momenta $\pi
^{a}{_{\mu }}$ can be obtained from the generating functional depending on
the old coordinates and the new momenta 
\begin{equation}
S_{X\pi }=-\int d^{p}\xi \ \partial _{a}X^{\mu }\pi ^{a}{_{\mu }}\ .
\end{equation}
One has 
\begin{eqnarray}
e_{a}{^{\mu }} &=&-\frac{\delta S_{X\pi }}{\delta \pi ^{a}{_{\mu }}}
=\partial _{a}X^{\mu } \ , \\
P_{\mu } &=&-\frac{\delta S_{X\pi }}{\delta X^{\mu }}=-\partial _{a}\pi ^{a}{
_{\mu }}\ .
\end{eqnarray}
Plugging this result into equation (\ref{s2}) one gets 
\begin{eqnarray}
S &=&\int d^{p+1}\xi \ \pi ^{a}{_{\mu }}\dot{e}_{a}{^{\mu }}+\phi ^{ab}{
_{\mu }}\partial _{\lbrack a}e_{b]}{^{\mu }}+  \nonumber \\
&&+\lambda ^{\mu }\left\{ \partial _{a}\pi ^{a}{_{\mu }}\pm T\frac{1}{p!}
\epsilon ^{a_{1}...a_{d}}\epsilon _{\mu \nu _{1}...\nu _{p}}e_{a_{1}}{^{\nu
_{1}}}...e_{a_{p}}{^{\nu _{p}}}\right\} \ ,  \nonumber \\
&&  \label{Lagr}
\end{eqnarray}
where $\phi ^{ab}{_{\mu }}$ are the Lagrange multiplier functions for the
constraints (\ref{noco}). In the case of a nonflat metric the Lagrangian 
(\ref{Lagr}) would be nonlocal in the new variables since it involves the
original coordinates $X^{\mu }$ present in the determinant of the metric $
G_{\mu \nu }$. But this problem does not arise in the case of a flat metric.
We have then the following action 
\begin{eqnarray}
S &=&\int d^{p+1}\xi \ \left[ \pi ^{a}{_{\mu }}\partial _{0}e_{a}{^{\mu }}
+\phi ^{ab}{_{\mu }}\partial _{\lbrack a}e_{b]}{^{\mu }}+\lambda ^{\mu
}\partial _{a}\pi ^{a}{_{\mu }}\pm \right.  \nonumber \\
&&\pm \left. T\lambda ^{\mu }\frac{1}{p!}\epsilon ^{a_{1}...a_{p}}\epsilon
_{\mu \nu _{1}...\nu _{p}}e_{a_{1}}{^{\nu _{1}}}...e_{a_{p}}{^{\nu _{p}}}
\right] \ ,  \label{s3}
\end{eqnarray}
which can be given in a covariant form if one identifies the Lagrange
multipliers $\lambda ^{\mu }$ with the time components of the tetrad fields, 
\begin{equation}
\lambda ^{\mu }=e_{0}{^{\mu }}\ ,
\end{equation}
and writes the momenta $\pi ^{a}{_{\mu }}$ and the Lagrange multipliers $%
\phi ^{ab}{_{\mu }}$ as the components of a $(p-1)$-form $F_{\mu }$, 
\begin{eqnarray}
\pi ^{a}{_{\mu }} &=&\epsilon ^{ab_{1}...b_{p-1}}F_{b_{1}...b_{p-1}\mu } \ ,\\
\phi ^{ab}{_{\mu }} &=&(p-1)\epsilon
^{abc_{1}...c_{p-2}}F_{0c_{1}...c_{p-2}\mu }\ .
\end{eqnarray}
Equation (\ref{s3}) then becomes 
\begin{eqnarray}
S &=&\int d^{p+1}\xi \ \left[ \epsilon ^{\alpha _{0}...\alpha _{p}}\partial
_{\alpha _{0}}e_{\alpha _{1}}{^{\mu }}F_{\alpha _{2}...\alpha _{p}\mu
}\right. \pm  \nonumber \\
&&\pm \left. T\frac{1}{(p+1)!}\epsilon ^{\alpha _{0}...\alpha _{p}}\epsilon
_{\nu _{0}...\nu _{p}}e_{\alpha _{0}}{^{\mu _{0}}}...e_{\alpha _{p}}{^{\mu
_{p}}}\right] \ ,
\end{eqnarray}
which can be compactly written in the differential form language as 
\begin{equation}
S=\int F_{\mu }\wedge de^{\mu }\pm T\frac{1}{(p+1)!}\epsilon _{\nu
_{0}...\nu _{p}}e^{\nu _{0}}\wedge ...\wedge e^{\nu _{p}}\ ,
\label{topact}
\end{equation}
where $e^{\nu }$ is a one-form and $F_{\mu }$ is a $(p-1)$-form. The action 
(\ref{topact}) is explicitly topological since it does not involve the metric.
After all one can see just at level of actions that the actions 
(\ref{linact}), (\ref{topact}) are equivalent to each other. 
This equivalence can be
established by integrating out the field $F_{\mu }$.

Now let us take a look at the symmetries and equiations of motions of the
action (\ref{topact}). The action has the following obvious symmetry 
\begin{equation}
\delta F_{\mu }=dw_{\mu } \ ,
\end{equation}
which is the shift $F_{\mu }$ by any exact $(p-1)$-form. There is one extra
symmetry which is less obvious 
\begin{eqnarray}
\delta e^{\mu } &=&df^{\mu } \ ,  \label{symm2} \\
\delta F_{\mu } &=&\pm \frac{T}{(p-1)!}\epsilon _{\mu \nu _{1}\nu _{2}...\nu
_{p}}f^{\nu _{1}}e^{\nu _{2}}\wedge ...\wedge e^{\nu _{p}}\ ,
\end{eqnarray}
where $f^{\mu }$ is an arbitrary zero-form (function). The equations of
motion are the following 
\begin{eqnarray}
de^{\mu } &=&0 \ ,  \label{eqmot} \\
dF_{\mu } &=&\mp \frac{T}{p!}\epsilon _{\mu \nu _{1}...\nu _{p}}e^{\nu
_{1}}\wedge ...\wedge e^{\nu _{p}}\ .
\end{eqnarray}
The classical moduli space is given by gauge non-equivalent solutions of
equations (\ref{eqmot}). Thus we were able to reformulate the topological $p$%
-brane theory in a locally flat background as an abelian BF-like model \cite
{Horowitz:1989ng} with the action given by (\ref{topact}). The model has a
bunch of U(1) fields $e^{\mu }$ and the nontriviality comes from the last
''mass'' term which mixes different gauge fields. For the case $p\geq 2$ the
action (\ref{topact}) can be thought as the zero gravitational constant
limit for the general relativity with cosmological constant in $(p+1)$
dimensional space-time. This limit should be taken in the first order
formalism \cite{Sa}.

The degrees of freedom (the classical moduli space) for the action 
 (\ref{topact}) can be analyzed in a straightforward fashion through the
cohomology groups. In general the situation depends on the details of the
topology of the background manifold or more precisely, on the structure of
the first cohomology group $H^{1}({\cal M},R)$. Since the one-forms $e^{\mu
} $ are closed and any two solutions that differ by an exact one-form are
gauge equivalent, $e^{\mu }$ is a element of $H^{1}({\cal M},R)$. The
equations for $F_{\mu }$ are more difficult to analyze since the right hand
side involves $e^{\mu }$. If $dim(H^{1}({\cal M},R))<p$ then the last
equation of motion in (\ref{eqmot}) reduces to $dF_{\mu }=0$. We have not
enough elements of the first cohomology group to construct a non-zero
right-hand side. Thus in this case the model coincides with $(p+1)$ copies
of an abelian BF system \cite{Horowitz:1989ng}. Therefore the space of
solutions for $e^{\mu }$ and $F_{\mu }$ is given by $p+1$ copies of $H^{1}(%
{\cal M},R)\oplus H^{p-1}({\cal M},R)$. In the case ${\cal M}=R\times \Sigma 
$ we have 
\begin{equation}
H^{1}({\cal M},R)\approx H^{1}(\Sigma ,R)\approx H^{p-1}(\Sigma ,R)\approx
H^{p-1}({\cal M},R)
\end{equation}
where we used Poincar\'{e} duality on the $p$-dimensional manifold $\Sigma $%
. Thus the space of gauge inequivalent solutions is even dimensional and it
is given by the product of $2(p+1)$ copies of the first cohomology group: $%
H^{1}(\Sigma ,R)$. The situation with $dim(H^{1}({\cal M},R))\geq p$ is more
involved. One should analyze what kind of right hand side in the last
equation (\ref{eqmot}) can be constructed from $e^{\mu }$. For instance in
the case ${\cal M}=R\times \Sigma $ it might be possible to construct out of 
$e^{\mu }$ the volume form for $\Sigma $: $e^{1}\wedge ...\wedge e^{p}$.
Since the volume form cannot be exact the corresponding equation has no
solution. We will not analyze this situation in all generality. However the
task might be solved straightforwardly as soon as we know explicitly the
content of $H^{1}({\cal M},R)$. Above analysis of degrees of 
 freedom is appropriate for the actions (\ref{linact})
 and (\ref{topact}) where the degenerate solutions are included.
  However to incorporate into the analysis 
 the restriction of nondegeneracy can be hard since the 
 removal of degennerate solutions from the phase space might
 destroy the gauge orbits.   
 The similar problem appears in the relation
 between $2+1$ gravity and Chern-Simons theory \cite{Matschull:1999he}.

On a curved space-time manifold there is no such simple BF-like action as in
locally flat case. However one can write the following action 
\begin{equation}
S =\int (dX^{\mu }-\eta ^{\mu })\wedge B_{\mu }\pm 
T \frac{1}{(p+1)!}
\sqrt{-G}\epsilon _{\mu _{0}...\mu _{p}}\eta ^{\mu
_{0}}\wedge ...\wedge \eta ^{\mu _{p}},  \label{BX}
\end{equation}
which is classically equivalent to the Nambu-Goto action (\ref{linact}). In
the action (\ref{BX}) $\eta ^{\mu }$ and $B_{\mu }$ are $1$-forms and $p$
-forms respectively. The action is nonlinear in $X^{\mu }$ and therefore it
is difficult to analyze it in the same fashion as before. The case $p=1$ is
definitly special. By itself the Nambu-Goto action (\ref{linact}) can be
interpreted as topological sigma model \cite{Witten:1988xj} in two
dimensions since $\sqrt{-G}\epsilon _{\mu \nu }$ might serve as closed
symplectic form on ${\cal M}$. Also the Nambu-Goto action is equivalent to
the following action 
\begin{equation}
S=\int dX^{\mu }\wedge \eta _{\mu }\pm \frac{T}{2}(-{\cal G})^{-1/2}\epsilon
^{\mu \nu }\eta _{\mu }\wedge \eta _{\nu }
\end{equation}
which is the Poisson sigma model on two dimensional ${\cal M}$ \cite
{Ikeda:1994fh}. Therefore we see that two dimensional topological string
theory is classically equivalent to other known theories up to some 
 subtleties related to degenerate configurations.

\section{Discussion and outline}

In the present work we considered the classical aspects of closed $p$-brane
theory defined on $(p+1)$ dimensional background manifolds ${\cal M}$. We
analyzed the hamiltonian and BRST structure of the theory. We saw that model
has different equivalent realizations. However the classical equivalence
between the constraints and the actions might fail at the quantum level due
to normal ordering problem (different regularizations). One can look at the
most familiar example $p=1$. For the case of quadratic constraints there is
an anomaly in the Virasoro algebra and therefore the system is not first
class anymore. In the case of linear constraints (\ref{s2}) there is no
anomaly possible since the constraints are completely linear. This
discussion gives us an example that at the quantum level the Nambu-Goto
action (natural source for linear constraints) and the Polyakov action (the
natural source for quadratic constraints) are not equivalent to each other.
 As well at the classical level different status of degenerate solutions 
 can bring extra problems into identification of two theories.

The actions (\ref{linact}) and (\ref{BX}) have a straightforward
  generalization to
the following topological models 
\begin{equation}
S=T\frac{1}{(p+1)!} \int d^{p+1}\xi C_{\mu _{0}...\mu _{p}}(X)
\epsilon^{\alpha _{0}...\alpha _{p}} \partial _{\alpha _{0}}
X^{\mu _{0}}...\partial _{\alpha _{p}}X^{\mu _{p}}
\end{equation}
and 
\begin{equation}
S =\int (dX^{\mu }-\eta ^{\mu })\wedge B_{\mu }
+T\frac{1}{(p+1)!}C_{\mu _{0}...\mu _{p}}(X)\eta ^{\mu _{0}}\wedge
...\wedge \eta ^{\mu _{p}},
\end{equation}
where $C$ is a $(p+1)$-form defined on the $d$-dimensional background
manifold ${\cal M}$ ($d$ might be any value equal or greater than $(p+1)$).
If the form $C$ is closed the model has many similarities with the
topological $p$-brane studied in the present work. We shall consider the
classical and quantum aspects of these theories in coming work.

\bigskip \begin{flushleft} {\Large\bf Acknowledgments} \end{flushleft}

\noindent We are grateful to Ansar Fayyazuddin for discussions.

\end{document}